\documentclass[%
 aip,
 amsmath,amssymb,
 reprint,%
]{revtex4-1}

\usepackage{graphicx}
\usepackage{dcolumn}
\usepackage{bm}

\usepackage[utf8]{inputenc}
\usepackage[T1]{fontenc}
\usepackage{mathptmx}

\begin{document}

\preprint{AIP/123-QED}

\title{A method to measure superconducting transition temperature of microwave kinetic inductance detector by changing power of readout microwaves}

\author{H. Kutsuma}
 \email{h.kutsuma@astr.tohoku.ac.jp}
 \affiliation{Astronomical institute, Tohoku University, 6-3 Aramaki, Aoba-ku, Sendai 980-8578, Japan}
 \affiliation{RIKEN Center for Advanced Photonics, 2-1 Hirosawa, Wako 351-0198, Japan}
 
 \author{Y. Sueno}
\affiliation{Kyoto University, Kitashirakawa-Oiwakecho, Sakyo-ku, Kyoto 606-8502, Japan}
 
\author{M. Hattori}%
\affiliation{Astronomical institute, Tohoku University, 6-3 Aramaki, Aoba-ku, Sendai 980-8578, Japan}

\author{S. Mima}%
\affiliation{RIKEN Center for Advanced Photonics, 2-1 Hirosawa, Wako 351-0198, Japan}

\author{S. Oguri}
\affiliation{Kyoto University, Kitashirakawa-Oiwakecho, Sakyo-ku, Kyoto 606-8502, Japan}

\author{C. Otani}%
\affiliation{RIKEN Center for Advanced Photonics, 2-1 Hirosawa, Wako 351-0198, Japan}

\author{J. Suzuki}
\affiliation{Kyoto University, Kitashirakawa-Oiwakecho, Sakyo-ku, Kyoto 606-8502, Japan}

\author{O. Tajima}
\affiliation{Kyoto University, Kitashirakawa-Oiwakecho, Sakyo-ku, Kyoto 606-8502, Japan}

\date{\today}

\begin{abstract}
A microwave kinetic inductance detector (MKID) is a cutting-edge superconducting detector, and its principle is based on a superconducting resonator circuit. The superconducting transition temperature ($T_c$) of the MKID is an important parameter because various MKID characterization parameters depend on it. In this paper, we propose a method to measure the $T_c$ of the MKID by changing the applied power of the readout microwaves. A small fraction of the readout power is deposited in the MKID, and the number of quasiparticles in the MKID increases with this power. Furthermore, the quasiparticle lifetime decreases with the number of quasiparticles. Therefore, we can measure the relation between the quasiparticle lifetime and the detector response by rapidly varying the readout power. From this relation, we estimate the intrinsic quasiparticle lifetime. This lifetime is theoretically modeled by $T_c$, the physical temperature of the MKID device, and other known parameters. We obtain $T_c$ by comparing the measured lifetime with that acquired using the theoretical model. Using an MKID fabricated with aluminum, we demonstrate this method at a $0.3~\mathrm{K}$ operation. The results are consistent with those obtained by $T_c$ measured by monitoring the transmittance of the readout microwaves with the variation in the device temperature. The method proposed in this paper is applicable to other types, such as a hybrid-type MKID.
\end{abstract}

\maketitle

A superconducting detector is a sensitive device because its gap energy is much smaller than that of a semiconductor detector. A microwave kinetic inductance detector (MKID) \cite{day2003broadband} is a superconducting microwave resonator. This resonant circuit is fabricated of a thin superconductor film on a silicon or sapphire substrate. The resonant frequency is tuned by the total length of the resonator. Therefore, numerous detectors can be read by multi-frequency tones using the same readout line. This advantage allows the production of a large MKID array, such as one containing 100--1000 detectors per single readout line. This has resulted in the rapid progress of radio and infrared astronomical observations \cite{endo2019first, monfardini2010nika, galitzki2014next, nagasaki2018groundbird}.

An MKID has a simple detection mechanism. Radiation entering the detector breaks a Cooper pair in the resonator when the radiation energy is larger than twice the gap energy of the superconducting film. A broken Cooper pair yields two additional quasiparticles. The inductance of the MKID circuit varies with the number of quasiparticles, and the resonant condition of this MKID changes with the inductance. We can measure entering radiation energy using the variation in the resonant phase and amplitude. 

The superconducting transition temperature ($T_c$) of the MKID is an important parameter. This is because various MKID characterization parameters depend on $T_c$. In this paper, we propose a method to measure the $T_c$ of the MKID by changing the readout power rapidly.

According to the BCS theory \cite{bardeen1957theory}, the relation between the gap energy ($\Delta$), and $T_c$ is given by following formula:
\begin{equation}
\label{eq:delta}
    \Delta = 1.76 k_\mathrm{B}T_c,
\end{equation}
where $k_\mathrm{B}$ is the Boltzmann constant. Using $\Delta$, the number of quasiparticles ($N_\mathrm{qp}$) under a low-temperature condition (i.e. $T\ll T_c$, where $T$ is the device temperature) is obtained by the following formula \cite{bardeen1957theory}: 
\begin{equation}
    N_\mathrm{qp} = 2N_0V\sqrt{2\pi k_BT\Delta}\exp\left(-\frac{\Delta}{k_\mathrm{B}T}\right),
\end{equation}
where $N_0$ is the single-spin density of states at the Fermi level ($N_0 = 1.74\times10^{10}~\mathrm{eV^{-1}\mu m^{-3}}$ for an aluminum \cite{mcmillan1968transition}), and $V$ is the volume of the resonator. The intrinsic quasiparticle lifetime, ($\tau^\mathrm{i}_\mathrm{qp}$), is obtained by the following formula \cite{kaplan1976quasiparticle}:
\begin{equation}
\label{eq:tau}
    \tau^\mathrm{i}_\mathrm{qp} = \frac{\tau_0}{\sqrt{\pi}}\left(\frac{k_\mathrm{B}T_c}{2\Delta}\right)^{5/2}\sqrt{\frac{T_c}{T}}\exp\left(\frac{\Delta}{k_\mathrm{B}T}\right),
\end{equation}
where $\tau_0$ is the electron--phonon interaction time ($\tau_0 = 458 \pm 10~\mathrm{ns}$ for an aluminum MKID\cite{de2011number}). The noise equivalent power derived from the generation and recombination of the quasiparticles ($NEP_\mathrm{gr}$) is obtained by the following formula\cite{sergeev2002ultrasensitive}:
\begin{equation}
    NEP_\mathrm{gr} = \frac{2\Delta}{\eta_\mathrm{pb}}\sqrt{\frac{N_\mathrm{qp}}{\tau^{\mathrm{i}}_\mathrm{qp}}},
\end{equation}
where $\eta_\mathrm{pb}$ is the pair breaking efficiency \cite{kurakado1982possibility} ($\eta_\mathrm{pb} = 0.57$ for aluminum\cite{kozorezov2000quasiparticle}).  
\begin{figure}
\includegraphics[width=7.8cm]{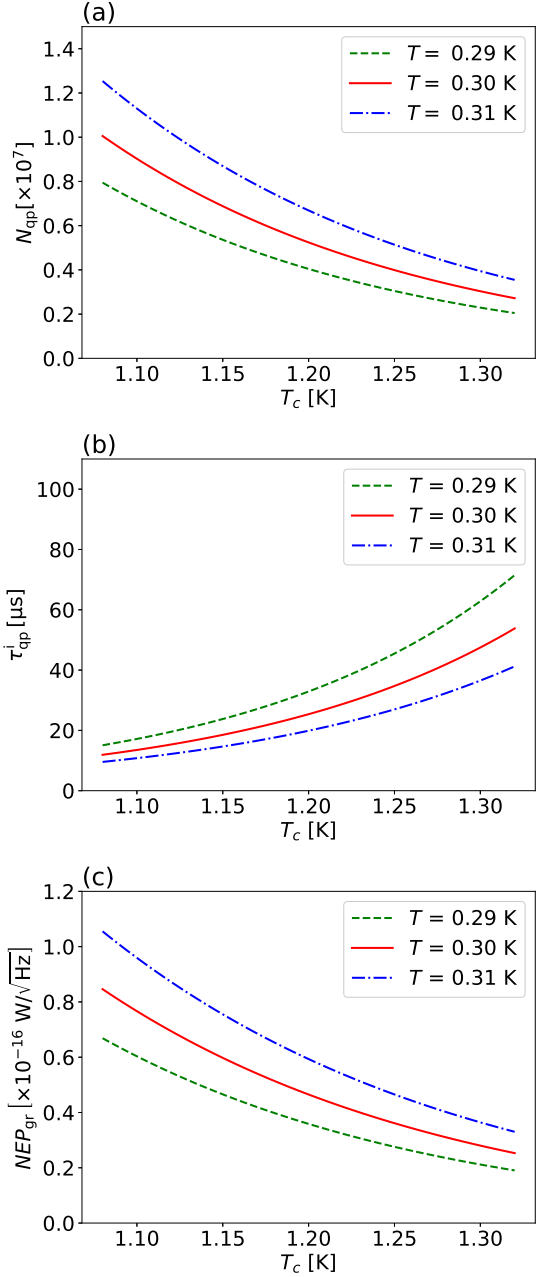}
\caption{\label{fig:temp_dep} Parameters characterizing an MKID as a function of the superconducting transition temperature, (a) the number of quasiparticles, (b) intrinsic quasiparticle lifetime, and (c) noise equivalent power. They are simulated considering an aluminum MKID with volume $1000~\mathrm{\mu m}^3$.}
\end{figure}
Figure \ref{fig:temp_dep} presents the parameters of an aluminum MKID as a function of $T_c$, where we display the plots of the device temperature at approximately $0.3~\mathrm{K}$. They are sensitive to $T_c$. Various previous studies have used different $T_c$, with the deviation being approximately $10\%$\cite{de2011number, fyhrie2020decay, vissers2020ultrastable}. Understanding the reason of this difference is a recent research topic, e.g., Fyhrie $\it{et~al.}$\cite{fyhrie2020decay} discusses the relation between $T_c$ and film thickness. 

Monitoring the transmittance of the readout microwaves with the device temperature variation is a conventional method to measure the $T_c$ of an MKID\cite{mazin2005, dominjon2016study}. This method is referred as the "S21 method" in this paper. Note that this method is not applicable for a hybrid-type MKID \cite{noroozian2009two, yates2011photon}.

Another method estimates $T_c$ using the power spectrum density ($S_x$), which is modeled by the following formula \cite{pieter2014}:
\begin{equation}
\label{eq:psd}
    S_x = \frac{4N_\mathrm{qp}\tau_\mathrm{qp}}{(1 + (2\pi f\tau_\mathrm{qp})^2)(1 + (2\pi f\tau_\mathrm{res})^2)}\left(\frac{\mathrm{d}x}{\mathrm{d}N_\mathrm{qp}}\right)^2 + X_\mathrm{system},
\end{equation}
where subscript $x$ denotes the phase or amplitude response, $\tau_\mathrm{qp}$ is the quasiparticle lifetime under the measurement condition, $f$ is the frequency of the detector response and $X_\mathrm{system}$ is the noise of the readout system. $\tau_\mathrm{res}$ is the resonator ring time given by $\tau_\mathrm{res} = Q_r/\pi f_r$ (where $Q_r$ is the quality factor of the resonance and $f_r$ is the resonant frequency). We extract $\tau_\mathrm{qp}$ by fitting the power spectrum density with the above formula. Under the assumption of $\tau^\mathrm{i}_\mathrm{qp} = \tau_\mathrm{qp}$, we obtain $T_c$ using Eq. (\ref{eq:tau}). This method requires that $X_\mathrm{system}$ is lower than the contribution of the first term in Eq (\ref{eq:psd}). Moreover, another contribution due to a two level system noise\cite{gao2007noise} should be low enough in the case of the phase response. 

We propose the third method to measure $T_c$, which uses a loss of the readout microwaves in the MKID. The quasiparticle lifetime ($\tau_\mathrm{qp}$) decreases with the increase of the number of additional quasiparticles ($N'_\mathrm{qp}$) produced by the readout power loss \cite{pieter2014, thompson2013dynamical, de2012microwave, de2014evidence, zmuidzinas2004superconducting}:
\begin{equation}
\label{eq:tau2}
    \tau_\mathrm{qp} = \frac{\tau^\mathrm{i}_\mathrm{qp}}{1 + \frac{N'_\mathrm{qp}}{N_\mathrm{qp}}}.
\end{equation}
Because the phase response ($\theta$) of the MKID is proportional to the number of additional quasiparticles \cite{gao2008equivalence}, the above formula is rewritten by:
\begin{equation}
\label{eq:tau3}
        \tau_\mathrm{qp} = \frac{\tau^\mathrm{i}_\mathrm{qp}}{1 + \alpha\theta},
\end{equation}
where $\alpha\theta = N'_\mathrm{qp}/N_\mathrm{qp}$. This suggests that $\tau^\mathrm{i}_\mathrm{qp}$ can be estimated from the relation between $\tau_\mathrm{qp}$ and $\theta$\cite{zmuidzinas2004superconducting, gao2008equivalence}. This relation is easily measured using our previous method to measure the phase responsivity\cite{kutsuma2019measurement}. A small fraction of the readout power is deposited in the MKID, and the response of the MKID increases with this power. In this paper, we demonstrate the measurement of this relation for an aluminum MKID. Subsequently, we obtain $T_c$ with Eq. (\ref{eq:tau}) and estimate $\tau^\mathrm{i}_\mathrm{qp}$.

\begin{figure}
\includegraphics[width=7cm]{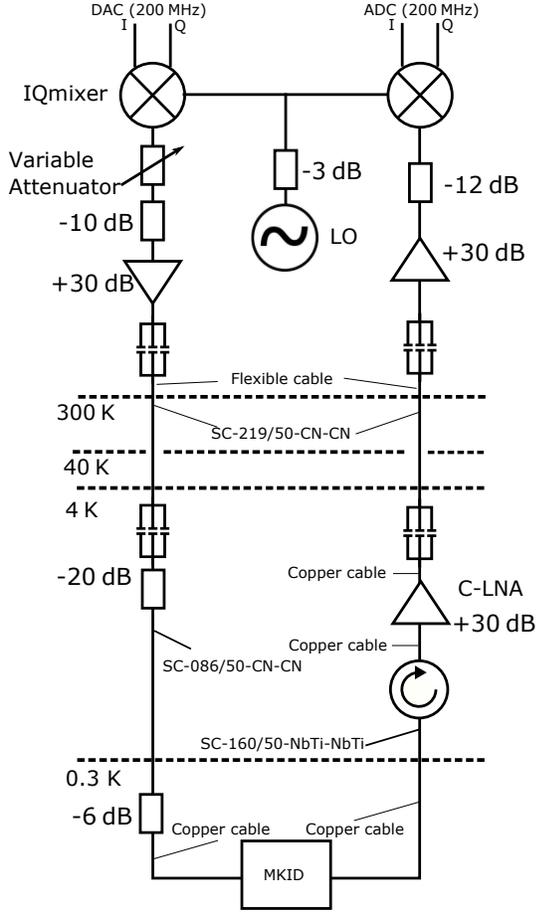}
\caption{\label{fig:setup}Diagram of the readout chain of the MKID. Our system feeds the readout signal at $200~\mathrm{MHz}$ bandwidth, which is up-converted into the microwave range. We use an NI Microwave Components FSL-0010 as a local oscillator (LO). The mixer is a Marki Microwave MLIQ-0218L. The input power into the MKID is controlled by a variable attenuator. The output microwaves from the MKID are amplified by a low noise amplifier (C-LNA, LNF-LN4$\_$8C, LOWNOISE FACTORY) and a warm amplifier (ZVE-8G+, Mini-Circuits). Subsequently, the down-converted signal is sampled.}
\end{figure}

Figure \ref{fig:setup} presents the diagram of the MKID readout.
Our cryostat (Niki Glass Co., Ltd.) consists of $4~\mathrm{K}$, and $40~\mathrm{K}$ thermal shields from inside to outside. They are insulated from the room temperature ($300~\mathrm{K}$) in a vacuum chamber and are cooled using a pulse tube refrigerator (PT407RM, Cryomech Co. LTD). The $4~\mathrm{K}$ thermal shield also acts as a magnetic shield (A4K, Amuneal Co. LTD) for mitigating the effects of geomagnetism. The MKID device is set in a light-tight copper box. The box is cooled by a $^3\mathrm{He}$-sorption refrigerator, and it is maintained at $T = 311~\mathrm{mK}$ with an accuracy of $6~\mathrm{mK}$.

The MKID device is fabricated at RIKEN. This device consists of a quarter-wave coplanar waveguide resonator and a feedline (there is no coupling with any antenna). The width of the center strip and gap of the resonator (the feedline) are $4~\mathrm{\mu m}$ ($12~\mathrm{\mu m}$)  and  $1.5~\mathrm{\mu m}$ ($8~\mathrm{\mu m}$), respectively. All the circuits patterns are formed using an aluminum film on a silicon wafer. The volume of the resonator is $2,600~\mathrm{\mu m}^3$ (the width is $4~\mathrm{\mu m}$, the length is $6,500~\mathrm{\mu m}$, and the thickness is $100~\mathrm{nm}$). Its resonant frequency and quality factor are $f_r = 4.30~\mathrm{GHz}$ and $Q_r = 2.61\times10^4$, respectively. Our readout system \cite{ishitsuka2016front, suzuki2018development} measures the response based on a direct down-conversion logic with a $200~\mathrm{MHz}$ sampling speed, and the data are down-sampled to a $1~\mathrm{MHz}$ step. The power of the readout microwave is controlled by a variable attenuator (LDA-602, Vaunix Co. LTD). It requires a few microseconds to change the attenuation value. 

\begin{figure}
\includegraphics[width=8cm]{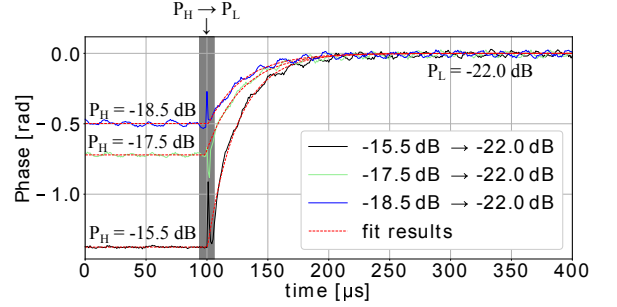}
\caption{\label{fig:switch} Phase responses to the rapid change in the power of the readout signal. The change is introduced at $t = 100~\mathrm{\mu s}$ with a few $\mathrm{\mu s}$ uncertainty. Therefore, we mask the region at $t = 95 \sim 105~\mathrm{\mu s}$ in the fitting. The dashed lines are the fitting results.}
\end{figure}

\begin{figure}
\includegraphics[width=8cm]{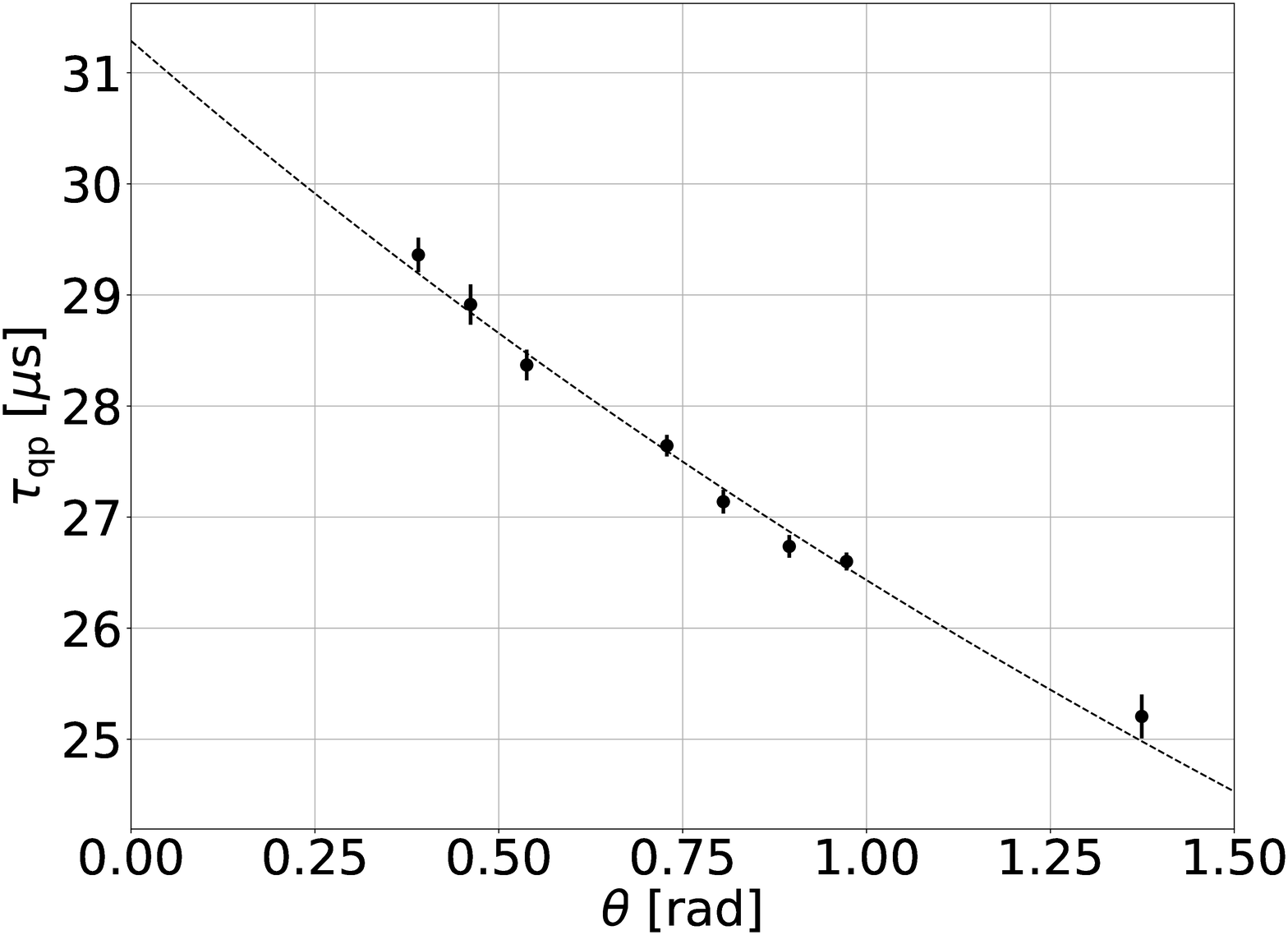}
\caption{\label{fig:Alresult}Measured relation between the quasiparticle lifetime ($\tau_\mathrm{qp}$) and the phase response ($\theta$). The black dots shows the measurement results. The dashed line represents the fit result. The value at $\theta=0$ corresponds to the intrinsic quasiparticle lifetime.}
\end{figure}

We use eight attenuation setups to change the readout power from high power ($P_\mathrm{H}$) to low power ($P_\mathrm{L}$), as summarized in Table \ref{tab:result}.  The readout power into the feedline is approximately $-65~\mathrm{dBm}$ at $P_\mathrm{L}$. We use the same treatment for the effects of a cable delay and linearity correction as described in the reference \onlinecite{kutsuma2019measurement}. Figure \ref{fig:switch} shows the phase response as a function of time. 
\begin{table}
\caption{\label{tab:result}Measured results for each setup of the readout power change. Only statistical errors are assigned here.}
\begin{ruledtabular}
\begin{tabular}{ccc}
$P_\mathrm{H}\rightarrow P_\mathrm{L} ~[\mathrm{dB}]$&$\theta~ [\mathrm{rad}]$&$\tau_\mathrm{qp}\mathrm[{\mu s}]$\\
\hline
$-15.5 \rightarrow -22.0$ & $1.374\pm 0.007$& $25.5\pm0.2$ \\
$-16.0 \rightarrow -22.0$ & $0.972\pm 0.004$&$26.6\pm0.1$ \\
$-16.5 \rightarrow -22.0$ & $0.894\pm 0.002$ &$26.7\pm0.1$ \\
$-17.0 \rightarrow -22.0$ & $0.805\pm 0.003$ &$27.1\pm0.1$ \\
$-17.5 \rightarrow -22.0$ & $0.728\pm 0.004$ &$27.6\pm0.1$ \\
$-18.0 \rightarrow -22.0$ & $0.538\pm 0.004$ &$28.4\pm0.1$ \\
$-18.5 \rightarrow -22.0$ & $0.462\pm 0.005$ &$28.9\pm0.2$ \\
$-19.0 \rightarrow -22.0$ & $0.391\pm 0.004$ &$29.4\pm0.2$ \\
\end{tabular}
\end{ruledtabular}
\end{table}
We reset the attenuation value at $t = 100~\mathrm{\mu s}$. We fit the data to Eq. (5) in the reference \onlinecite{kutsuma2019measurement}, and we obtain the phase response ($\theta$) and the quasiparticle lifetime ($\tau_\mathrm{qp}$) for each setup using Eq.(\ref{eq:tau3}). We mask the data in the short period, $t = 95-105~\mathrm{\mu s}$, because of the uncontrolled state of attenuation soon after the reset. We measure 50 samples for each set of power change. Table \ref{tab:result} summarizes the fitted results for each setup.
Figure \ref{fig:Alresult} displays the relation between $\tau_\mathrm{qp}$ and $\theta$. We obtain $\tau^\mathrm{i}_\mathrm{qp} = 31.3\pm0.2~\mathrm{\mu s}$ from the fit with Eq. (\ref{eq:tau3}). 
Subsequently, we obtain $T_c = 1.278\pm0.001~\mathrm{K}$ using Eq. (\ref{eq:tau}), where we only estimate the statistical error. We estimate the systematic uncertainties for $T_c$; device temperature ($0.025~\mathrm{K}$), time for changing the attenuation value ($0.014~\mathrm{K}$), and electron-phonon interaction time ($0.004~\mathrm{K}$).  We finally obtain $T_c = 1.28\pm0.03~\mathrm{K}$, including the systematic error.

\begin{figure}
\includegraphics[width=8cm]{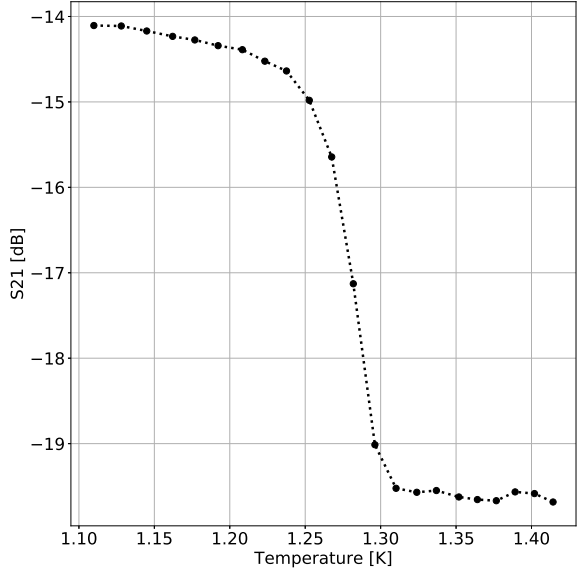}
\caption{\label{fig:S21_result} Transmittance of the readout microwaves as a function of the MKID temperature.}
\end{figure}

For comparison, we also measure $T_c$ by the S21 method. Figure \ref{fig:S21_result} presents the S21 intensity results at $4.35~\mathrm{GHz}$ as a function of the device temperature. We determine $T_c$ as a temperature at the middle of the transition, $T_c = 1.27\pm0.04~\mathrm{K}$. Here, the error includes the difference from the onset of the superconducting transition and the uncertainty of the thermometer. We confirm the consistency in the results from the two methods.

In summary, we propose a method to measure the $T_c$ of an MKID device by changing the power of the readout microwaves. In this method, we obtain the intrinsic quasiparticle lifetime using the device temperature. Subsequently, we estimate $T_c$ using them. We demonstrate our method using an aluminum MKID maintained at $311~\mathrm{mK}$. We obtain $T_c = 1.28\pm0.03~\mathrm{K}$. This result is consistent with the conventional method: monitoring the microwave transmittance by changing the device temperature, $T_c = 1.27\pm0.04~\mathrm{K}$. The method proposed in this paper is applicable to other types, such as a hybrid-type MKID.

This work was supported by the JRA program in RIKEN and Grants-in-Aid for Scientific Research from The Ministry of Education, Culture, Sports, Science and Technology, Japan (KAKENHI Grants 19H05499, 15H05743, 16J09435, and R2804). We thank Prof. Koji Ishidoshiro for lending us the cryostat used in this paper. We acknowledge Prof. Masato Naruse for the MKID design. We acknowledge Mr. Noboru Furukawa for the MKID fabrication. We thank Dr. Shunsuke Honda for useful discussions. OT acknowledges support from the Heiwa-Nakajima Foundation, US-Japan Science Technology Cooperation Program, and SPIRITS grant. 

The data that support the findings of this study are available from the corresponding author upon reasonable request.

\nocite{*}
\bibliography{aipsamp2}

\end{document}